\documentclass[preprints,article,accept,moreauthors,pdftex]{Definitions/mdpi} 

\usepackage[para, online, flushleft]{threeparttable}

\firstpage{1} 
\pubvolume{xx}
\issuenum{1}
\articlenumber{5}
\pubyear{2020}
\copyrightyear{2020}
\history{Received: date; Accepted: date; Published: date}

\Title{Debiased/Double Machine Learning for Instrumental Variable Quantile Regressions}

\Author{Jau-er Chen $^{1, 2, *}$, Chien-Hsun Huang$^{3}$  and Jia-Jyun Tien$^{4}$}

\AuthorNames{Jau-er Chen, Chien-Hsun Huang  and Jia-Jyun Tien}

\address{%
$^{1}$ \quad Institute for International Strategy, Tokyo International University. 
	1-13-1 Matobakita Kawagoe, Saitama 350-1197,  Japan.\\
$^{2}$ \quad Center for Research in Econometric Theory and Applications, National Taiwan University. 
	No. 1, Section 4, Roosevelt Road, Taipei 10617, Taiwan.\\
$^{3}$ \quad The Office of the Chief Economist, Microsoft Research. 
	Microsoft Building 99, 14820 NE 36th Street, Redmond, Washington 98052, USA. \\
$^{4}$ \quad Department of Economics, National Taiwan University. 
	No. 1, Section 4, Roosevelt Road, Taipei 10617, Taiwan.}

\corres{Correspondence: jechen@tiu.ac.jp; Tel.: +81-49-232-1111}

\abstract{
In this study, we investigate estimation and inference on a low-dimensional causal parameter 
in the presence of high-dimensional controls in an instrumental variable quantile regression. 
Our proposed econometric procedure builds on the Neyman-type orthogonal moment conditions of a previous study \cite{ref-Chernozhukovetal2018} and is thus relatively insensitive to the estimation of the nuisance parameters.
The Monte Carlo experiments show that the estimator copes well with high-dimensional controls. 
We also apply the procedure to empirically reinvestigate 
the quantile treatment effect of 401(k) participation on accumulated wealth.
}


\begin{document}

\section{Introduction}

Machine learning methods have been actively studied in economic big data settings in recent years, 
cf.\ \cite{ref-Athey2017} and \cite{ref-AtheyImbens2019}. 
Most empirical studies in economics aim to understand the program evaluation, or equivalently, the causal effect. Constructing the counterfactual and then estimating causal effects relies on an appropriately chosen identification strategy. In economics, the instrumental variable approach is an extensively used identification strategy for causal inference. Therefore, the machine learning techniques often require the adaptation to exploit the structure of the underlying identification strategy. These adaptations are part of  an emerging research area at the intersection of machine learning and econometrics, which is called the causal machine learning in the economic literature.
Two popular causal machine learning approaches are currently available to estimate treatment effects through adapted machine learning algorithms, and they also provide valid standard errors of an estimated causal parameter of interest, such as the average treatment effect and quantile treatment effect. These two approaches are the double machine learning (DML) cf.\ \cite{ref-Chernozhukov2018}, and the generalized random forests (GRF) of \cite{ref-Athey2019}. The GRF estimates heterogeneous treatment effects and explores variable importance accounting for heterogeneity in the treatment effect. The resulting information is crucial for optimal polices mapping from individuals' observed characteristics to treatments. The DML provides a clever and general recipe for use of sample splitting, cross-fitting, and Neyman orthogonalization, to make causal inference possible and allows for almost any machine learner. Furthermore, the DML is feasible for dealing with high-dimensional datasets where researchers observe massive characteristics of the units. For instance, through sample splitting, the DML estimates each of the nuisance function (e.g., the expectations for the target variable and outcome variable given high-dimensional controls) on an auxiliary sample, and then it uses out-of-sample residuals as the basis for the treatment effect estimation. Moreover, the cross-fitting algorithm allows researchers to use all of the data in the final treatment effect estimation instead of throwing an auxiliary sample used in the sample splitting earlier. This procedure in fact follows the Neyman-type orthogonal moment conditions which ensure that the estimation above is insensitive to the first-order perturbations of the nuisance parameter near the true value, and consequently, the regular inference on a low-dimensional causal parameter proceeds.

With the identification strategy of selection on observables (aka.\ unconfoundedness), 
empirical applications have been investigated by using the aforementioned two approaches, 
including the works by \cite{ref-Gilchrist2016} and \cite{ref-Davis2017}. 
When it comes to the identification strategy of selection on unobservables, 
few empirical papers that use  causal machine learning can be found in the existing literature. 
Those empirical applications very often lack important observed control variables or involve reverse causality,
and thus,  researchers resort to the instrumental variable approach.
In this study, we investigate the estimation and inference of a low-dimensional causal parameter in the presence of high-dimensional controls in an instrumental variable quantile regression. In particular, we build on a previous study \cite{ref-Chernozhukovetal2018} and then further concretize the econometric procedure. To the best of our knowledge, this study is the first to investigate Monte Carlo performance and empirical studies based on the DML procedure within the framework of instrumental variable quantile regressions. We also make our R codes available on the GitHub repository\footnote{The R scripts conducting the estimation and inference of the Double Machine Learning for Instrumental Variable Quantile Regressions  can be found at https://github.com/FieldTien/DML-IVQR/tree/master/example} so that other researchers can benefit from the proposed estimation method.

\cite{ref-ChenHsiang2019} investigated the instrumental variable quantile regression in the context of GRF. 
Their econometric procedure yielded a measure of variable importance 
in terms of characterizing heterogeneity in the treatment effect. 
They proceeded by empirically investigating the distributional effect of 401(k) participation on net financial assets. They demonstrated that income, age, education, 
and family size are the first four important variables in explaining treatment effect heterogeneity. 
In contrast to our study, their GRF-based estimator is not designed for high-dimensional settings. 
With the same dataset, 
we also apply the proposed procedure to empirically investigate the
distributional effects of the 401(k) participation on net financial assets. 
Empirical results signify that 
the 401(k) participants with low savings propensity are more associated with the nonlinear income effect, 
which complements the findings in studies conducted by \cite{ref-Chernozhukov2018} and \cite{ref-ChiouChenChen2018}. 
In addition, nonlinear transformations of the four aforementioned variables are also identified as important variables 
in the current context of DML-based instrumental variable quantile regression 
with high-dimensional observed characteristics.

The rest of the paper is organized as follows. 
The model specification and practical algorithm are introduced in Section 2, which includes detailed descriptions of a general recipe for the DML. 
Section 3 presents finite-sample performances of the estimator through Monte Carlo experiments. 
Section 4 reinvestigates an empirical study on quantile treatment effects: the effect of 401(k) participation on wealth.
Section 5 concludes the paper.

\section{The Model and Algorithm}

In this study, we use the instrumental variable quantile regression (IVQR) of \cite{ref-ChernozhukovHansen2005} and \cite{ref-ChernozhukovHansen2008} to identify the quantile treatment effect.
In Section 2.1, we briefly review the DML procedure developed in \cite{ref-Chernozhukov2018}.
In Section 2.2, we briefly review the conventional IVQR based on the exposition in \cite{ref-ChernozhukovHansen2005}. 
In Section 2.3, we present DML-IVQR within the framework of high-dimensional controls.

\subsection{The Double Machine Learning}

In this section, we briefly review the plain DML procedure.
Let us consider the following canonical example of estimating treatment effect $\alpha_0$ in a partial linear regression under the identification strategy of selection on observables.
\begin{align}
Y = D\alpha_0 + h_0(X) + U,   \quad \mathbb{E}[U|X, D]=0
\end{align}
where $Y$ is the outcome variable, $D$ is the target variable, and $X$ is a high-dimensional vector of controls.
$X$ are control variables in the sense that
\begin{align}
D= m_0(X) + V,
\end{align}
where $m_0(\cdot)\neq 0$ and $\mathbb{E}[V|X]=0$.
Note that $h_0(X)$ and $m_0(X)$ are nuisance functions because they are not the primary objects of interest.
\cite{ref-Chernozhukov2018} develop the DML procedure for estimating $\alpha_0$, which is outlined in the following three steps.
\begin{itemize}
\item[I.] [Sample splitting] \ Split the data into $K$ random and roughly equally sized folds. For $k=1, \ldots, K$,  a machine learner is used to fit the high-dimensional nuisance functions, $\hat{\mathbb{E}}_{(-k)}[Y|X]$ and $\hat{\mathbb{E}}_{(-k)}[D|X]$, using all data except for the $k$th fold.
\item[II.] [Cross-fitting and residualizing] \ Calculate out-of-sample residuals for these fitted nuisance functions on the $k$th fold; that is, $\hat{Y}_{(k)} = Y_{(k)} - \hat{\mathbb{E}}_{(-k)}[Y|X]$ and $\hat{D}_{(k)} = D_{(k)} - \hat{\mathbb{E}}_{(-k)}[D|X]$.
\item[III.] [Treatment effect estimation and inference] \ Collect all of the out-of-sample residuals from the cross-fitting stage, and use the ordinary least squares to regress $\hat{Y}$ on $\hat{D}$ to obtain $\check{\alpha}$, the estimator of $\alpha_0$.
The resulting $\check{\alpha}$ estimate can be paired with heteroskedastic consistent standard errors to obtain a confidence interval for the treatment effect.
\end{itemize}

Because estimating the nuisance functions through machine learners induces regularization biases,
the cross-fitting step was used to refrain from its biasing the treatment effect estimate.
The procedure is random due to the sample splitting. 
Different researchers with the same data set but making different random splits will obtain distinct estimators. This randomness can be reduced by using a larger value of $K$, but this increases computation cost. $K\geq 10$ is recommended.
In fact, the DML procedure follows a unified approach in terms of moment conditions and the Neyman orthogonality condition, cf.\ \cite{ref-ChernozhukovHansenSpindler2015}.
In a nutshell, we seek to find moment conditions
\begin{align}
\mathbb{E}\left[g\left(Y, D, X, \alpha_0, \eta_0 \right) \right] = 0
\end{align}
such that the following Neyman orthogonality condition holds
\begin{align}
\left.\partial_{\eta}\mathbb{E}\left[g\left(Y, D, X, \alpha_0, \eta_0 \right) \right]\right|_{\eta=\eta_0} = 0,
\end{align}
where $\eta_0$ are nuisance functions with the true values.
Equation (4) is insensitive to the first-order perturbations of the nuisance function $\eta$ near the true value. This property allows the estimation of  $\eta_0$ using regularized estimators (machine learners) $\hat{\eta}$. Without this property, regularization may have too much effect on the estimator of $\alpha_0$ for regular inference to proceed. 
The estimator $\check{\alpha}$ of $\alpha_0$ solves the empirical analog of the equation (3):
\begin{align*}
\frac{1}{n}\sum_{i=1}^n g\left(y_i, d_i, x_i, \check{\alpha}, \hat{\eta} \right)  = 0,
\end{align*}
where we have plugged in the estimator $\hat{\eta}$ for the nuisance function. 
Owing to the Neyman orthogonality property, the estimator is first-order equivalent to the infeasible estimator 
$\tilde{\alpha}$ solving
\begin{align*}
\frac{1}{n}\sum_{i=1}^n g\left(y_i, d_i, x_i, \tilde{\alpha}, \eta_0 \right)  = 0,
\end{align*}
where we use the true value of $\eta$. 

Therefore, we recast the canonical example set by equations (1) and (2) into the moment conditions that guide the DML procedure outlined above.
\begin{align*}
\mathbb{E}\left[g\left(Y, D, X, \alpha_0, \eta_0 \right) \right] 
&=\mathbb{E}\big{[} \left[ \left(Y-E[Y|X] \right) - \left(D-E[D|X] \right )\alpha_0 \right] \times  \left(D-E[D|X] \right)\big{]}\\
&=\mathbb{E}\big{[} \left[\left(D\alpha_0 + h_0(X) + U - \left(m_0(X)\alpha_0 + h_0(X)\right)  \right) -V\alpha_0 \right] \times V\big{]}\\
&=\mathbb{E}\left[ \left(D\alpha_0 - m_0(X)\alpha_0 + U -V\alpha_0 \right)\times V \right]\\
&=\mathbb{E}\left[ m_0(X)\alpha_0 V + V^2\alpha_0 - m_0(X)\alpha_0 + UV - V^2\alpha_0 \right]\\
&=\mathbb{E}\left[  UV \right] = 0,
\end{align*}
where $\eta_0 = \left[ \mathbb{E}[Y|X] \ \ \mathbb{E}[D|X] \right]$.
It is easy to see that the corresponding Neyman orthogonality condition holds 
\begin{align*}
&\left.\partial_{\eta}\mathbb{E}\left[g\left(Y, D, X, \alpha_0, \eta_0 \right) \right]\right|_{\eta=\left[ \mathbb{E}[Y|X] \ \ \mathbb{E}[D|X] \right]} \\
& \qquad= \left.\partial_{\eta}\mathbb{E}\big{[} \left[ \left(Y-E[Y|X] \right) - \left(D-E[D|X] \right )\alpha_0 \right] \times  \left(D-E[D|X] \right)\big{]}    \right|_{\eta=\left[ \mathbb{E}[Y|X] \ \ \mathbb{E}[D|X] \right]} =0.
\end{align*}

\subsection{The Instrumental Variable Quantile Regression}

Based on the exposition in \cite{ref-ChernozhukovHansen2005}, the following conditional moment restriction yields an IVQR estimator:
\begin{equation}
\mathbb{P}[Y \leq q(\tau,D,X)|X,Z ]= \tau,
\end{equation}
where $q(\cdot)$ is the structural quantile function, $\tau$ is the quantile index,
Y is the outcome variable, D is the target (endogenous) variable,
and $X$ and $Z$ are control variables and instruments, respectively.
Equation (1) and linear structural quantile specification lead to the following unconditional moment restriction
\begin{equation}
\mathbb{E}[(\tau - \mathbf{1}(Y - D'\alpha - X'\beta \leq 0)\Psi]=0
\end{equation}
where $$\Psi:= \Psi(X, Z)$$ is a vector of the  function of the instruments and control variables,
and $(\alpha', \beta')'$ are the unknown parameters. In particular, $\alpha$ is a causal parameter of interest.
The parameters depend on the quantile of interest, 
but we suppress  $\tau$ associated with $\alpha$ and $\beta$ for simplicity of presentation.
Equation (2) leads to a particular moment condition for residualization. That is

\begin{equation}
g_{\tau}(\alpha;\beta,\delta) = \left(\tau - \mathbf{1}(Y \leq D'\alpha + X'\beta)\right)\Psi(\alpha,\delta(\alpha))
\end{equation}
with the instrument
\begin{equation}\Psi(\alpha,\delta(\alpha)) := (Z - \delta(\alpha)X)\end{equation}
\begin{equation*}\delta(\alpha) = M(\alpha)J^{-1}(\alpha), \end{equation*}
where $\delta$ is a matrix parameter for weighting the least square $Z$ on the $X$ coefficient,
\begin{equation*}
M(\alpha) = \mathbb{E}[ZX'f_{\varepsilon}(0|X,Z)],\quad J(\alpha) = \mathbb{E}[XX'f_{\varepsilon}(0|X,Z)]
\end{equation*}
and $f_{\varepsilon}(0|X,Z)$ is the conditional density of $\epsilon = Y - D'\alpha - X'\beta(\alpha)$
with $\beta(\alpha)$  defined by
\begin{equation}
\mathbb{E}[(\tau - \mathbf{1}(Y \leq D'\alpha + X'\beta(\alpha))X]=0.
\end{equation}

First, we construct the grid search interval for $\alpha$  and then profile out the coefficient for each $\alpha$ in the interval on the exogenous variable using equation (5). Specifically,
\begin{equation*}
\hat{\beta}(a) = \arg \min_{b \in \mathcal{B}} \frac{1}{N}\sum_{i=1}^N \rho_{\tau}(Y_i - D_{i}'a - X_{i}'b).
\end{equation*}
By substituting these  estimates into the sample counterpart of the moment restriction (2), we obtain
\begin{equation}
\hat{g}_{N}(a) = \frac{1}{N}\sum_{i=1}^N g(a,\hat{\beta}(a),\hat{\delta}(a)),
\end{equation}
where
\begin{equation*}
\hat{\delta}(a) = \widehat{M}(a)\widehat{J}^{-1}(a)
\end{equation*}
with
\begin{equation*}
\widehat{M}(a) = \frac{1}{Nh_N}\sum_{i=1}^N Z_i X_i'K_{h_N}\big(Y_i - D_i'a-X_i'\hat{\beta}(a)\big)
\end{equation*}
\begin{equation*}
\widehat{J}(a) = \frac{1}{Nh_N}\sum_{i=1}^N X_i X_i'K_{h_N}\big(Y_i - D_i'a-X_i'\hat{\beta}(a)\big)
\end{equation*}
where $K_{h_N}$ is a kernel function with bandwidth $h_N$.
In the Monte Carlo simulations, we assume that we know the density function according to our data generation process. Thus, we can solve for the parameters by optimizing the criterion function of generalized method of moments (GMM) as follows: 
\begin{equation}
\hat{\alpha}(\tau) = \arg \min_{a \in \mathcal{A}} N\hat{g}_N(a)'\widehat{\Sigma}(a,a)^{-1}\hat{g}_N(a),
\end{equation}
where
\begin{equation*}
\widehat{\Sigma}(a_1,a_2) = \frac{1}{N}\sum_{i=1}^N g\big(a_1,\hat{\beta}(a_1)\big)g\big(a_2,\hat{\beta}(a_2)\big)'
\end{equation*}
is a weighting matrix used in the GMM estimation.
Note that the estimator $\hat{\alpha}$ based on the inverse quantile regression (i.e., IVQR) of \cite{ref-ChernozhukovHansen2008} is the first-order equivalent to the estimator defined by the GMM above.

\subsection{Estimation with High-dimensional Controls}

We modify the procedure presented in Subsection 2.2  to deal with a dataset of high-dimensional control variables. 
To this end, we construct the grid search interval for $\alpha$ and profile the coefficients on exogenous variables using the $L_1$-norm penalized quantile regression estimator of \cite{ref-BelloniChernozhukov2011}:
\begin{equation}
\hat{\beta}(a) = \arg \min_{b \in \mathcal{B}} \frac{1}{N}\sum_{i=1}^N \rho_{\tau}(Y_i - D_i'a - X_i'b) + \lambda\sum_{j=1}^{dim(b)} \hat{\sigma}_j
|b_j|,
\end{equation}
where $\rho(\cdot)$ is the check function and $\hat{\sigma}^2_j = (1/n)\sum_{i=1}^n x^2_{ij}$.
The penalty level $\lambda$ is chosen as follows.
\begin{equation}
\lambda = 2 \cdot \Lambda(1-\alpha | X),
\end{equation}
where $\Lambda(1-\alpha | X):= (1-\alpha)$-quantile of $\Lambda$ conditional on $X$.
The random variable 
\begin{equation}
\Lambda = n \sup_{u\in \mathcal{U}} \max_{1\leq j\leq dim(b)}
\left| \frac{1}{n}\sum_{i=1}^n \left[ \frac{x_{ij}(u - \mathbb{I}\{u_i \leq u\} )}{\hat{\sigma}_j \sqrt{u(1-u)} } \right] \right|,
\end{equation}
where $u_1, \ldots, u_n$ are i.i.d.\ uniform (0, 1) random variables that are independently distributed from the controls $x_1, \ldots, x_n$. The random variable $\Lambda$ has a pivotal distribution conditional on 
$X = [x_1, \ldots, x_n]'$. Therefore, we compute $\Lambda(1-\alpha | X)$ using simulation of $\Lambda$.
\cite{ref-BelloniChernozhukov2011} show that the aforementioned choice for the penalty level $\lambda$ leads to the optimal rates of convergence for the $L_1$-norm penalized quantile regression estimator.
Namely, 
the choice of the penalization parameter $\lambda$ based on (13) is theoretically grounded and feasible.
In high-dimensions setting, $K$-fold cross-validation is very popular in practice.
However, computational cost is roughly proportional to $K$.
The recently derived non-asymptotic error bounds in \cite{ref-Chetverikovetal2021} imply that the $K$-fold cross-validated  Lasso estimator has nearly optimal convergence rates. 
Although their theoretical guarantees do not directly apply to the $L_1$-norm penalized quantile regression estimator,  it still sheds some light on the use of cross-validation as an alternative to determine the penalty level $\lambda$ in our analysis.\footnote{We have conducted Monte Carlo experiments indicating that the choice of $\lambda$ based on (13) or 5-fold cross-validation leads to similar finite sample performances of our proposed procedure in terms of root-mean-square error, mean absolute error, and bias.  Simulation findings are tabulated in Section 3. When there are many binary control variables, the $L_1$-norm penalized quantile regression may suffer singularity issues in estimation. If this is the case, researchers can utilize the algorithm developed by \cite{ref-YiHuang2017} using the Huber loss function to approximate the quantile loss function.}

In addition, we estimate
\begin{equation}
\widehat{M}(a) = \frac{1}{Nh_N}\sum_{i=1}^N Z_i X_i'K_{h_N}\big(Y_i - D_i'a-X_i'\hat{\beta}(a)\big)
\end{equation}
and
\begin{equation}
\widehat{J}(a) = \frac{1}{Nh_N}\sum_{i=1}^N X_i X_i'K_{h_N}\big(Y_i - D_i'a-X_i'\hat{\beta}(a)\big).
\end{equation}

We also perform dimension reduction on $J$ because of the large dimension of $X$.
In particular, we implement the following regularization.
\begin{equation}
\hat{\delta}_j(a) = \arg \min_{\delta} \frac{1}{2} \delta'\hat{J}(a)\delta - \hat{M}_j(a)\delta + \vartheta||\delta||_1.
\end{equation}
The regularization above does a weighting Lasso for each instrument variable on control variables.
Consequently, the $L_1$ norm optimization obeys the Karush-Kuhn-Tucker condition
\begin{equation*}
||\hat{\delta}_j(a)'\hat{J}(a) - \hat{M}_j(a)||_{\infty} \leq \vartheta, \quad \forall j.
\end{equation*}
More importantly, the aforementioned procedure is the double machine learning algorithm for the IVQR, which satisfies the Neyman orthogonality condition as follows.
Let us present the IVQR as a first-order-equivalent GMM estimator.
To this end, we define
\begin{equation*}
g(\alpha, \eta) = \left(\tau - \mathbf{1}(Y \leq D'\alpha + X'\beta)\right) \left( Z - \delta(\alpha) X \right)
\end{equation*}
where $\eta = [\beta(\alpha)' \   \delta(\alpha)' ]'$ are high-dimensional nuisance parameters in the DML setting discussed in Section 2.1 with true values 
$\eta_0 = [\beta(\alpha_0)' \   \delta(\alpha_0)' ]'$.
Therefore,
\begin{align}
\mathbb{E}\left[g(\alpha_0, \eta_0)\right] 
&=\mathbb{E}\left[\left(\tau - \mathbf{1}(Y \leq D'\alpha_0 + X'\beta_0)\right) \left( Z - \delta(\alpha_0) X \right)\right]\notag\\
&=\mathbb{E}\left[ \mathbb{E}\left[ \left. \tau - \mathbf{1}(Y \leq D'\alpha_0 + X'\beta_0) \right| X, Z \right] \left( Z - \delta(\alpha_0) X \right)\right]=0.
\end{align}
We then calculate
\begin{align*}
\left.\partial_{\eta} \mathbb{E}\left[g(\alpha_0, \eta)\right] \right|_{\eta=\eta_0}
= \left.\begin{array}{c} \left.\partial_{\beta} \mathbb{E}\left[g(\alpha_0, \eta)\right] \right|_{\eta=\eta_0} \\
\left.\partial_{\delta} \mathbb{E}\left[g(\alpha_0, \eta)\right] \right|_{\eta=\eta_0} \end{array}\right. .\\
\end{align*}
Specifically,
\begin{align}
\left.\partial_{\beta} \mathbb{E}\left[g(\alpha_0, \eta)\right] \right|_{\eta=\eta_0}
&= \partial_{\beta} \mathbb{E}\left[ \mathbb{E}\left[ \left. \tau - \mathbf{1}(Y \leq D'\alpha_0 + X'\beta_0) \right| X, Z \right] \left( Z - \delta(\alpha_0) X \right)\right]\\
&= \partial_{\beta} \mathbb{E}\left[ \left( \left. \tau - F(Y \leq D'\alpha_0 + X'\beta_0  \right| X, Z) \right) \left( Z - \delta(\alpha_0) X \right)\right]\notag\\
&= \mathbb{E}\left[ZX' f_{\epsilon}(0|X, Z) \right] - \delta(\alpha_0) \mathbb{E}\left[XX' f_{\epsilon}(0|X, Z) \right]\notag\\
&= M(\alpha_0) -\delta(\alpha_0) J(\alpha_0)\notag\\
&= M(\alpha_0) -M(\alpha_0)J^{-1}(\alpha_0) J(\alpha_0)=0.\notag
\end{align}
\begin{align}
\left.\partial_{\delta} \mathbb{E}\left[g(\alpha_0, \eta)\right] \right|_{\eta=\eta_0}
&= \partial_{\delta} \mathbb{E}\left[\left(\tau - \mathbf{1}(Y \leq D'\alpha_0 + X'\beta_0)\right) \left( Z - \delta(\alpha_0) X \right)\right]\\
&= - \mathbb{E}\left[\left(\tau - \mathbf{1}(Y \leq D'\alpha_0 + X'\beta_0)\right) X \right] =0.\notag
\end{align}
We thus verify that $\left.\partial_{\eta} \mathbb{E}\left[g(\alpha_0, \eta)\right] \right|_{\eta=\eta_0}=0$,
which indicates the Neyman orthogonality condition holds.

After implementing the DML outlined above,
we solve for the low-dimensional causal parameter $\alpha$ by optimizing the GMM defined as follows.
The sample counterpart of the moment condition
\begin{equation}
\hat{g}_N(a) = \frac{1}{N} \sum_{i=1}^N \big(\tau - \mathbf{1}\big(Y_i - D_i'a -X_i'\hat{\beta}(a) \leq 0\big)\big)\Psi(a,\hat{\delta}(a)).
\end{equation}
Accordingly,
\begin{equation}
\hat{\alpha} = \arg \min_{a \in \mathcal{A}} N\hat{g}_N(a)'\widehat{\Sigma}(a,a)^{-1}\hat{g}_N(a).
\end{equation}

\cite{ref-ChernozhukovHansenSpindler2015} show that the key condition enabling us to perform valid inference on $\alpha_0$ is the adaptivity condition:
$\sqrt{N}\left(\hat{g}(\alpha_0, \hat{\eta}) -  \hat{g}(\alpha_0, \eta_0) \right)
\stackrel{\mathbb{P}_N}{\longrightarrow} 0$.
In particular, each element $\hat{g}_j$ of $\hat{g}=\left(\hat{g}_j \right)_{j=1}^k$
can be expanded as
$\sqrt{N}\left(\hat{g}_j(\alpha_0, \hat{\eta}) -  \hat{g}_j(\alpha_0, \eta_0) \right)
=T_{1, j} + T_{2, j} +T_{3, j}$, which are formally defined on page 663 in their paper.
The term $T_{1, j}$ vanishes precisely because of orthogonality, that is, $T_{1, j}=0$.
However, the terms $T_{2, j}$ and $T_{3, j}$ do not vanish.
The $T_{2, j}$ and $T_{3, j}$ vanish when cross-fitting and sample splitting are implemented. 
These two terms are also asymptotically negligible when we impose a further structure on the problem:
such as using a sparsity-based machine learner (e.g., $L_1$-norm penalized quantile regression)
under approximate sparsity conditions. 
In our procedure, equations (12) and (17) are sparsity-based machine learners.
Therefore, we use no cross-fitting in the DML-IVQR algorithm.

Theoretically speaking, based on equation (19), the approach can be applied to machine learners other than the Lasso. The chief difficulty in implementing an estimation based on Equation (19) is that 
the function being minimized is both non-smooth and non-convex,
and any machine learners are used to dealing with a functional response variable in this context,
cf.\ \cite{ref-BCFH2017}.
In addition, the corresponding DML with non-linear equations is difficult. 
Therefore, our practical strategy is to implement the DML-IVQR procedure 
described in equations (12) -- (17) and (21) -- (22), which is equivalent to the Neyman orthogonality condition defined in (19) and (20).

\subsection{Weak-Identification Robust Inference}

Under the regularity conditions listed in \cite{ref-ChernozhukovHansen2008}, 
asymptotic normality for the GMM estimator with a non-smooth objective function is guaranteed.
We have
\begin{equation*}
\sqrt{n}\hat{g}_N(a) \stackrel{d} \longrightarrow  N(0,\Sigma(a,a)).
\end{equation*}
Consequently, it leads to
\begin{equation*}
N\hat{g}_N(a)'\widehat{\Sigma}(a,a)^{-1}\hat{g}_N(a) \stackrel{d} \longrightarrow  \chi^2_{dim(Z)}.
\end{equation*}
We define
$$
W_N\equiv N\hat{g}_N(a)'\widehat{\Sigma}(a,a)^{-1}\hat{g}_N(a).
$$
It  follows that a valid
$(1-p)$ percent confidence region for the true parameter, $\alpha_0$, can be constructed as the set
\begin{equation*}
CR := \{\alpha \in \mathcal{A}:W_N (\alpha) \leq c_{1-p}\},
\end{equation*}
where $c_{1-p}$ is the critical point such that
\begin{equation*}
P[\chi^2_{dim(Z)} > c_{1-p}]=p,
\end{equation*}
and $\mathcal{A}$ can be numerically approximated by the grid $\{\alpha_j,j=1,...,J\}$.

\section{Monte Carlo Experiments}

We evaluate the finite-sample performance, in terms of 
mean bias (BIAS), mean absolute error (MAE) and root-mean-square error (RMSE)
of the DML-IVQR through 1000 simulations.
The following data generating process is modified from that considered in \cite{ref-ChenLee2018}.

$$\begin{bmatrix}
    u_i \\
    \epsilon_i\\
    
\end{bmatrix}
\sim N\left(0,\begin{bmatrix}
    1  & 0.3 \\
    0.3  & 1\\
    
\end{bmatrix}\right)\\
\medskip
$$
$$\begin{bmatrix}
    x_{ji} \\
    z_{1i}\\
    z_{2i}\\
    v_{1i}\\
    v_{2i}\\ 
\end{bmatrix}
\sim N (0,I)\\
\medskip
$$
$$Z_{1i} = z_{1i} + x_{2i}+ x_{3i}+  x_{4i}+ v_{1i}$$
$$Z_{2i} = z_{2i} + x_{7i}+ x_{8i} + x_{9i}+ x_{10i}+ v_{2i}$$
$$D_i = \Phi(z_{1i}+z_{2i} +\epsilon_i) $$
$$X_{ji} = \Phi(x_{ji})$$
$$Y_i =1+ D_i + 5X_{1i} + 5X_{2i} + 5X_{3i} + 5X_{4i} + 5X_{5i} + 5X_{6i} + 5X_{7i} + D_i\times u_i,$$
where $\Phi(\cdot)$ is the cumulative distribution function of a Normal random variable;
$i=1, 2, \ldots, n$; $j= 1, 2, \ldots, p$; $p$ is the dimension of controls $X$, and $p=100$.
There are ten relevant controls:  $X_{1i}, \ldots, X_{10i}$. 
The instrumental variable is $Z$.
The target variable is $D$.
Consequently,
$$\alpha(\tau)=1+F_\epsilon^{-1}(\tau),$$
where $\tau$ is the quantile index and $F_\epsilon (\cdot)$ is the cumulative distribution function of the random variable $\epsilon$. 
Therefore, the median treatment effect $\alpha(0.5)=1$.

\subsection{Residualizing $Z$ on $X$}

We focus on comparing the BIAS, MAE and RMSE resulting from different procedures under the exact specification (10 control variables).  res-GMM represents residualizing $Z$ on $X$.
GMM stands for doing no residualizing $Z$ on $X$.
Table 1 shows that residualizing $Z$ on $X$ leads to an efficiency gain across quantiles especially when the sample size is moderate. 

\begin{table}[h]
  \caption{Residualizing and nonResidualizing Z on X}
  \begin{center}
  \begin{threeparttable}
    \begin{tabular}{lcccccc}
      \hline
           & \multicolumn{3}{c}{n = 500} & \multicolumn{3}{c}{n = 1000}\\
      \hline
            
          & \multicolumn{1}{c}{RMSE} & \multicolumn{1}{c}{MAE} & \multicolumn{1}{c}{BIAS} & \multicolumn{1}{c}{RMSE} & \multicolumn{1}{c}{MAE} &  \multicolumn{1}{c}{BIAS} \\
            \hline
      $\alpha_{0.10}$ (res-GMM)  & 0.1888 & 0.1510  & -0.0893 & 0.1219 & 0.0950 & -0.0551\\
      $\alpha_{0.10}$ (GMM) & 0.4963 & 0.2559 & -0.1775 & 0.1631 & 0.1138 & -0.0627 \\
      
      \hline
      $\alpha_{0.25}$ (res-GMM) & 0.1210   &  0.0966  &-0.0334  & 0.0812 & 0.0654 & -0.0256\\
      $\alpha_{0.25}$ (GMM) & 0.1782  & 0.1179 & -0.0254   & 0.0963 & 0.0754 & -0.0234\\
      \hline
      $\alpha_{0.50}$ (res-GMM)   & 0.0989   &  0.0716  &0.0091   & 0.0689 & 0.0436 & -0.0020\\
      $\alpha_{0.50}$ (GMM) & 0.1436  & 0.1016 &  0.0340  & 0.0801 & 0.0542 & 0.0078\\
      \hline
      $\alpha_{0.75}$ (res-GMM)  & 0.1374  & 0.1066  &0.0552  & 0.0828 & 0.0676 &0.0212\\
      $\alpha_{0.75}$ (GMM) & 0.2403 & 0.1710 & 0.1294 & 0.1146 & 0.0848 & 0.0442\\
      
      \hline
      $\alpha_{0.90}$ (res-GMM)  & 0.2437  &  0.1839 & 0.1225  & 0.1391 & 0.1067  &0.0667\\
      $\alpha_{0.90}$ (GMM) & 0.8483  & 0.5340 & 0.4959  & 0.3481 & 0.1967 & 0.1613\\
      \hline
      \end{tabular}%
    
    \label{table 1}%
    
\begin{tablenotes}
  \tiny
  The date generating process considers ten control variables. res-GMM represents residualizing  $Z$ on $X$. The GMM does not residualize  $Z$ on $X$. $\alpha_{\tau}$ denotes the quantile treatment effect.
  \end{tablenotes}
  \end{threeparttable}
  \end{center}
  \label{table:simDisimCoefNewDef}
  \end{table}

\subsection{IVQR with High-dimensional Controls}

We now evaluate the finite-sample performance of the IVQR with high-dimensional controls.
The data generating process involves 100 control variables with an approximate sparsity structure.
In particular, the exact model (true model) depends only on 10 relevant control variables out of the 100 controls. 
Let's fix the name of different estimators first.
The full-GMM uses 100 control variables without regularization. 
The oracle-GMM knows the identity of the true controls and then uses the ten relevant variables.
The DML-IVQR is our proposed estimator.
Table 2 shows that the RMSE  stemmed from the DML-IVQR are close to those from the oracle estimator. The numbers in parentheses are ratios of RMSE or MAE of any estimator to those of the oracle-GMM. 
The BIAS and MAE indeed signify that the DML-IVQR achieves a lower bias in the simulation study.  
In addition,  Figure 1 plots the distributions of the IVQR estimator with and without double machine learning.
The DML-IVQR stands for the double machine learning for the IVQR with high-dimensional controls.
Histograms signify that the DML-IVQR estimator is more efficient and less biased than IVQR using many control variables. Because a weak-identification robust inference results naturally from the IVQR, we construct the robust confidence regions for the full-GMM, oracle-GMM and the DML-IVQR estimators. 
In Figures 2-4, the vertical axis displays the value of the test statistic $W_N(\alpha)$ which is defined in Section 2.4. The horizontal line in gray is the 95\% critical value from $\chi^2_{dim(Z)}$.
\cite{ref-ChernozhukovHansen2008} robust confidence region is all values of $\alpha$ such that
the $W_N(\alpha)$ lies below the horizontal line. 
The robust inferential procedure is still valid when identification is weak or fails partially or completely.
Thus Figures 2-4 show that, across quantiles, the robust confidence region based on the DML-IVQR is relatively sharp compared to those of the full-GMM.
In addition, the confidence regions based on the DML-IVQR are remarkably close to those obtained by the oracle estimator.

As to the choice of penalty parameter, 
researchers can chose $\lambda$ based on Equation (13) proposed by \cite{ref-BelloniChernozhukov2011}  
or based on the $K$-fold cross-validation.
Both methods of choosing $\lambda$  lead to similar finite sample performances of DML-IVQR in terms of the RMSE, MAE and BIAS.  Simulation findings are summarized in Table 3.

\begin{table}
  \centering
  \caption{IVQR with High-dimensional Controls}

\par
  \medskip
    \begin{threeparttable}
    \begin{tabular}{lccccc}
    \hline
           & \multicolumn{2}{c}{n = 500} \\
    \hline
          & \multicolumn{1}{c}{RMSE (ratio)} & \multicolumn{1}{c}{MAE (ratio)} & \multicolumn{1}{c}{BIAS} \\
          \hline
    $\alpha_{0.10}$ (full-GMM)     & 0.7648 (4.05) & 0.6645 (4.40) & -0.6533 \\
    $\alpha_{0.10}$ (oracle-GMM)    & 0.1888 (1.00) & 0.1510 (1.00)  & -0.0893  \\
    $\alpha_{0.10}$ (DML-IVQR)      & 0.3112 (1.64) & 0.2389 (1.58) & -0.2039  \\
    \hline
    $\alpha_{0.25}$ (full-GMM)     & 0.2712 (2.24) & 0.2212 (2.28) &  -0.1876 \\
    $\alpha_{0.25}$ (oracle-GMM)    & 0.1210 (1.00)   &  0.0966 (1.00)  &-0.0334  \\
    $\alpha_{0.25}$ (DML-IVQR)       & 0.1562 (1.29) & 0.1254 (1.29) & -0.0796 \\
    \hline
    
    $\alpha_{0.50}$ (full-GMM)     & 0.1627 (1.64) & 0.1234 (1.72) & 0.0190 \\
    $\alpha_{0.50}$ (oracle-GMM)    & 0.0989 (1.00)   &  0.0716 (1.00)  &0.0091 \\
    $\alpha_{0.50}$ (DML-IVQR)       & 0.1168 (1.18) & 0.0846 (1.18) & -0.0186 \\
    \hline
     $\alpha_{0.75}$ (full-GMM)    & 0.3421 (2.48) & 0.2806 (2.63) & 0.2502\\
    $\alpha_{0.75}$ (oracle-GMM)    & 0.1374 (1.00)  & 0.1066 (1.00)  &0.0552  \\
    $\alpha_{0.75}$ (DML-IVQR)       & 0.1495 (1.08) & 0.1167 (1.09)  & 0.0516 \\
    \hline
    $\alpha_{0.90}$ (full-GMM)    & 0.9449 (3.87) & 0.8032 (4.36) & 0.7891 \\
    $\alpha_{0.90}$ (oracle-GMM)    & 0.2437 (1.00)   &  0.1839 (1.00)  & 0.1225  \\
    $\alpha_{0.90}$ (DML-IVQR)       & 0.3567 (1.46) & 0.2608 (1.41)& 0.2011 \\
    \hline
    \hline
           &  \multicolumn{2}{c}{n = 1000}\\
    \hline
          &  \multicolumn{1}{c}{RMSE (ratio)} & \multicolumn{1}{c}{MAE (ratio)}& \multicolumn{1}{c}{BIAS}\\
          \hline
    $\alpha_{0.10}$ (full-GMM)     & 0.3917 (3.21) & 0.3442 (3.62) & -0.3303  \\
    $\alpha_{0.10}$ (oracle-GMM)    & 0.1219 (1.00) & 0.0950 (1.00) & -0.0551 \\
    $\alpha_{0.10}$ (DML-IVQR)      & 0.1376 (1.12) & 0.1085 (1.14)  &-0.0759  \\
    \hline
    $\alpha_{0.25}$ (full-GMM)      & 0.1646 (2.02) & 0.1361 (2.08) &-0.1134 \\
    $\alpha_{0.25}$ (oracle-GMM)     & 0.0812 (1.00) & 0.0654 (1.00) & -0.0256 \\
    $\alpha_{0.25}$ (DML-IVQR)       & 0.0991 (1.22) & 0.0804 (1.22) &-0.0436 \\
    \hline
    
    $\alpha_{0.50}$ (full-GMM)      & 0.1038 (1.50) & 0.0754 (1.72) & -0.0002 \\
    $\alpha_{0.50}$ (oracle-GMM)     & 0.0689 (1.00) & 0.0436 (1.00) & -0.0020 \\
    $\alpha_{0.50}$ (DML-IVQR)       & 0.0775 (1.12) & 0.0510 (1.16)  &-0.0142 \\
    \hline
     $\alpha_{0.75}$ (full-GMM)     & 0.1747 (2.10) & 0.1452 (2.14) &0.1174  \\
    $\alpha_{0.75}$ (oracle-GMM)    & 0.0828 (1.00) & 0.0676 (1.00) &0.0212 \\
    $\alpha_{0.75}$ (DML-IVQR)       & 0.0930 (1.12) & 0.0741 (1.09)  &0.0226 \\
    \hline
    $\alpha_{0.90}$ (full-GMM)     & 0.4320 (3.10) & 0.3681 (3.45) & 0.3495  \\
    $\alpha_{0.90}$ (oracle-GMM)     & 0.1391 (1.00) & 0.1067 (1.00)  & 0.0667 \\
    $\alpha_{0.90}$ (DML-IVQR)        & 0.1649 (1.18) & 0.1231 (1.15) & 0.0731  \\
   \hline
    \end{tabular}%
    \begin{tablenotes}
    \tiny The full-GMM uses 100 control variables without regularization. The oracle-GMM uses the ten relevant variables.
DML-IVQR is a double machine learning procedure. $\alpha_{\tau}$ denotes the quantile treatment effect. The numbers in parentheses are the ratios of the RMSE or MAE  of any estimator to those of the oracle-GMM.
    \end{tablenotes}
    \end{threeparttable}

  \label{table 2}%
\end{table}

 \begin{figure}[htp]

  \centering

  \label{figur}\caption{Histograms of the DML-IVQR Estimates (in green)}
  \par
  
  \bigskip
  
  \begin{tabular}{cc}


    \includegraphics[width=68mm]{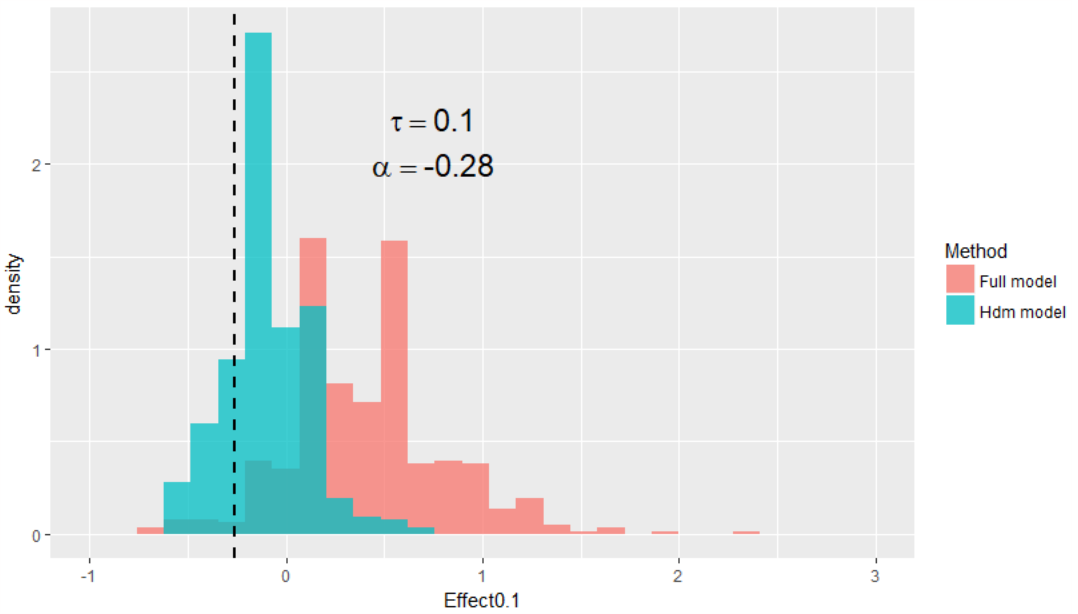}&

    \includegraphics[width=68mm]{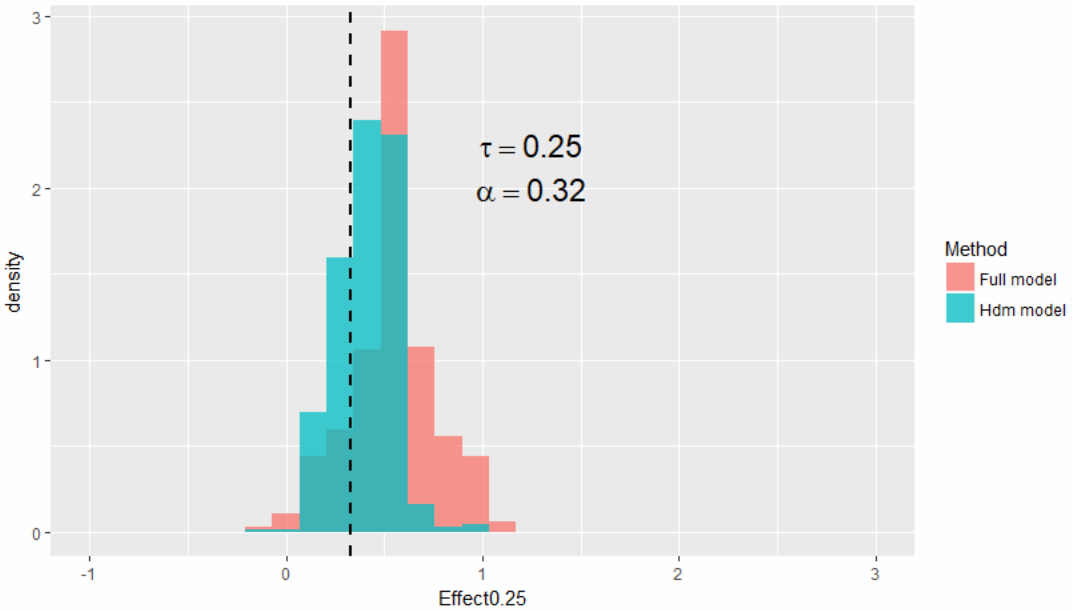}\\

    \includegraphics[width=68mm]{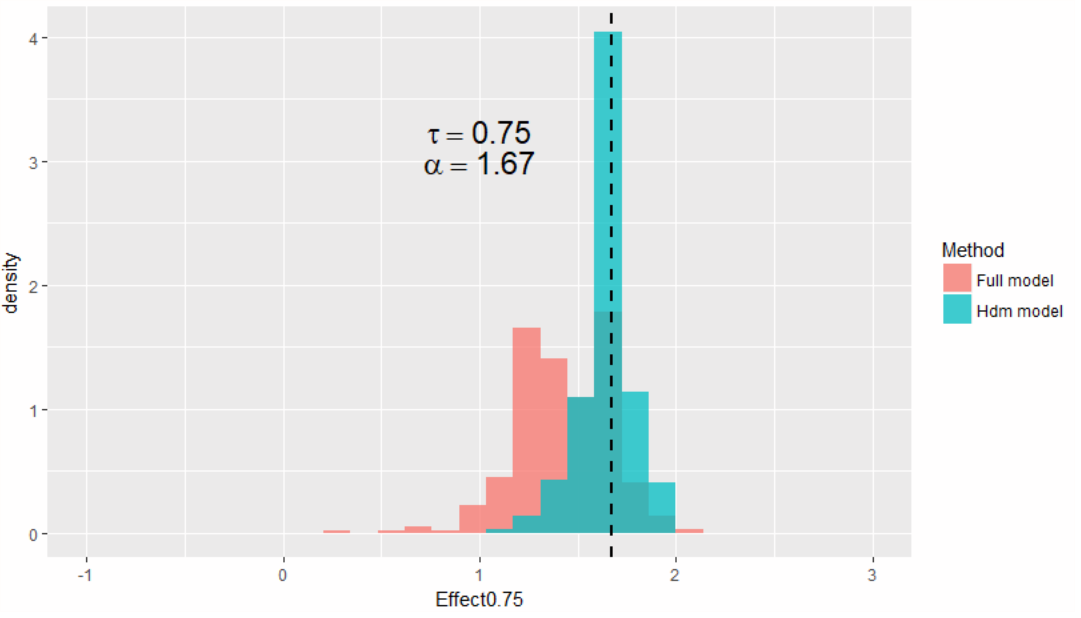}&

    \includegraphics[width=68mm]{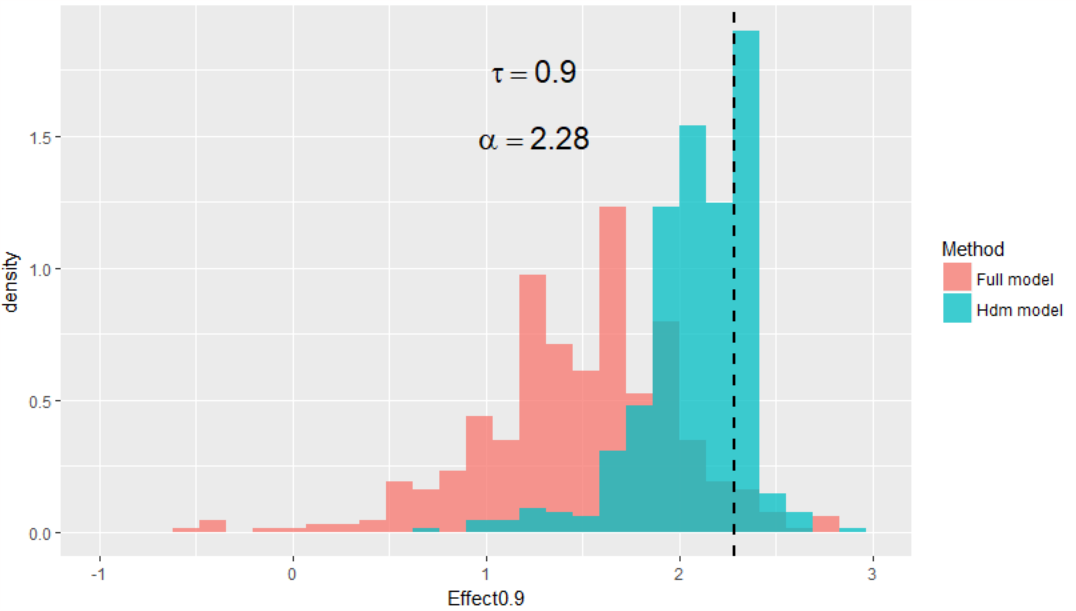}\\
    
  \end{tabular}
  \begin{center}
    \includegraphics[width=68mm]{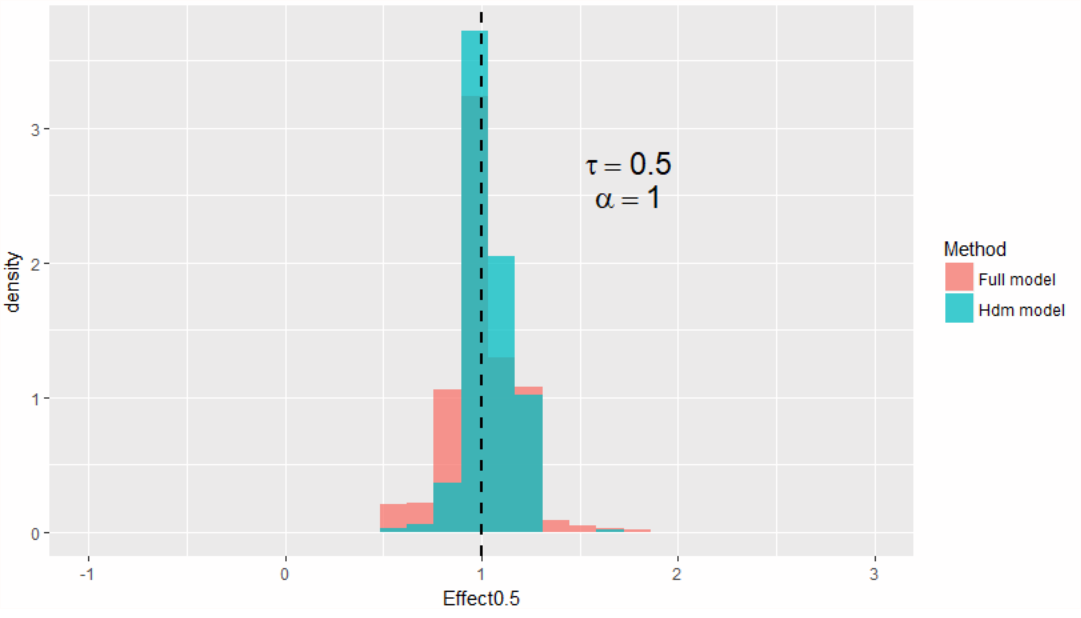}
  \end{center}
  \scriptsize
  
\end{figure}

\begin{figure}[htp]
\centering

 \label{figur}
 \caption{Weak-Instrument Robust Inference at $0.5$th quantile: \\ DML-IVQR (in brown), oracle-GMM, and full-GMM.}
\par

\bigskip
  \includegraphics[scale=0.33]{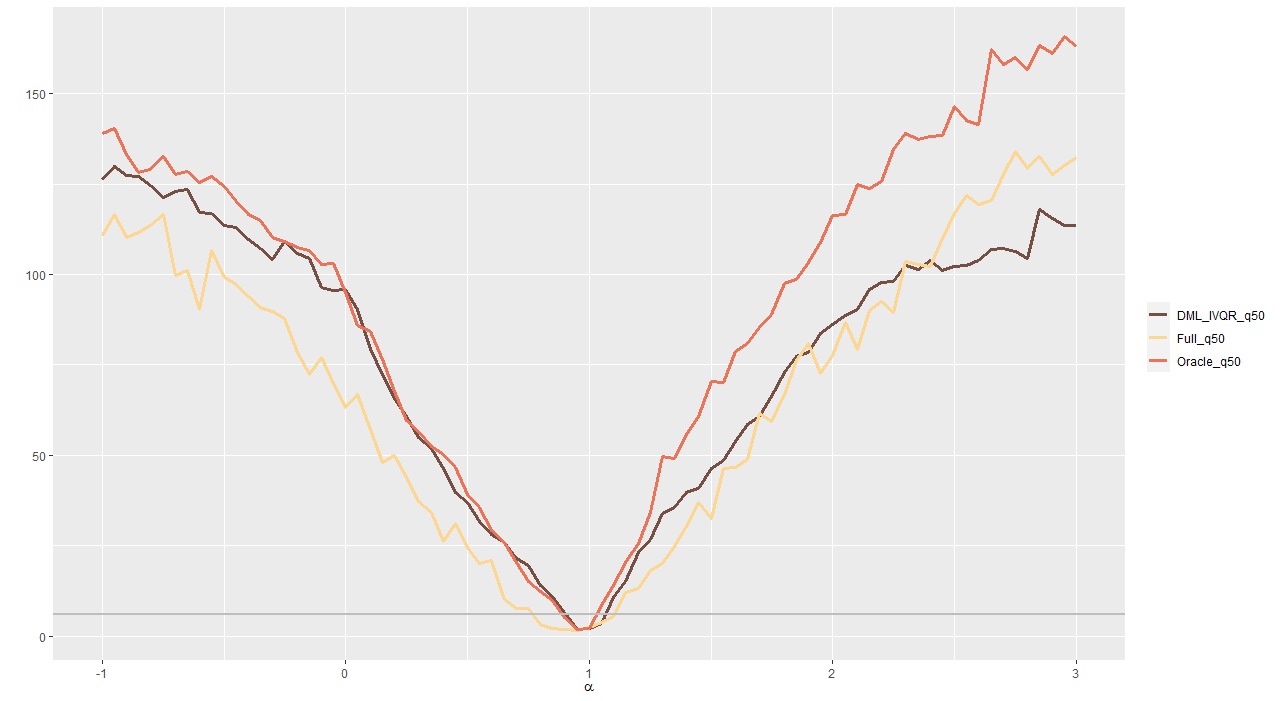}

\end{figure}

\begin{figure}[htp]

  \centering

  \label{figur}
  \caption{Weak-Instrument Robust Inference at $0.25$th quantile: \\ DML-IVQR (in brown), oracle-GMM, and full-GMM.}
\par

\bigskip
  \includegraphics[scale=0.33]{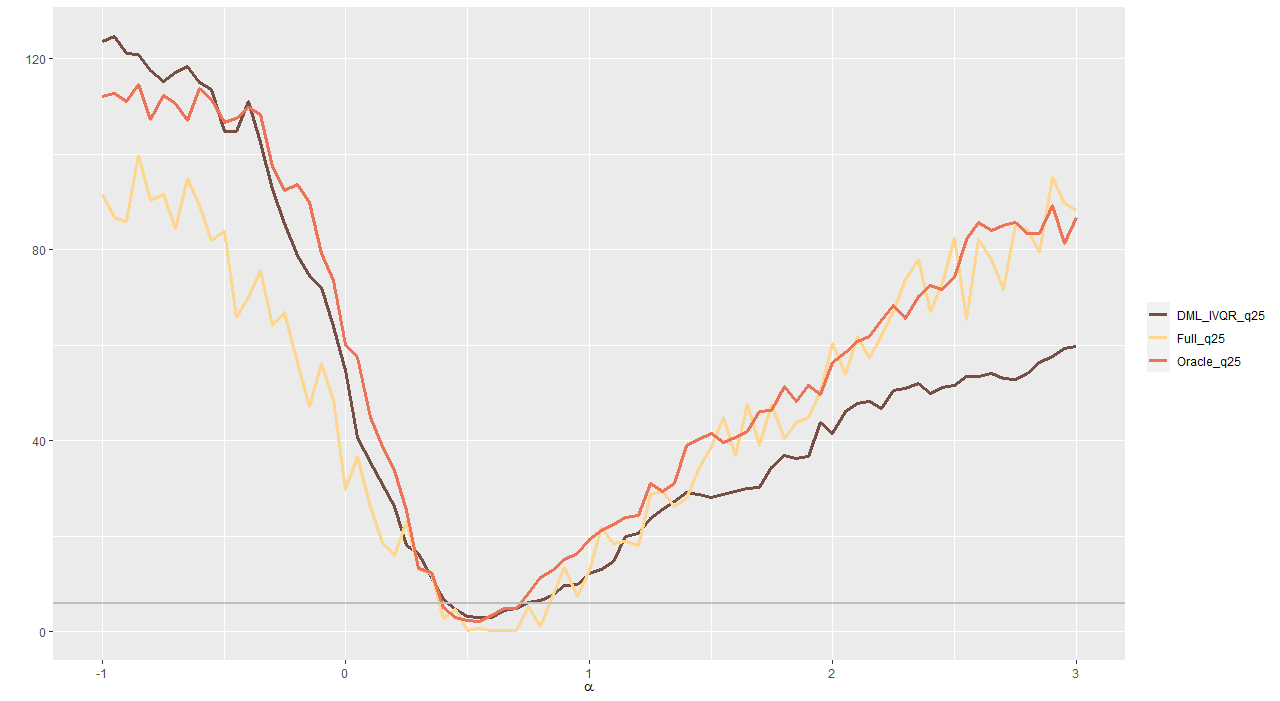}

\end{figure}

\begin{figure}[htp]

  \centering

  \label{figur}
 \caption{Weak-Instrument Robust Inference at $0.75$th quantile: \\ DML-IVQR (in brown), oracle-GMM, and full-GMM.}
\par

\bigskip
  \includegraphics[scale=0.33]{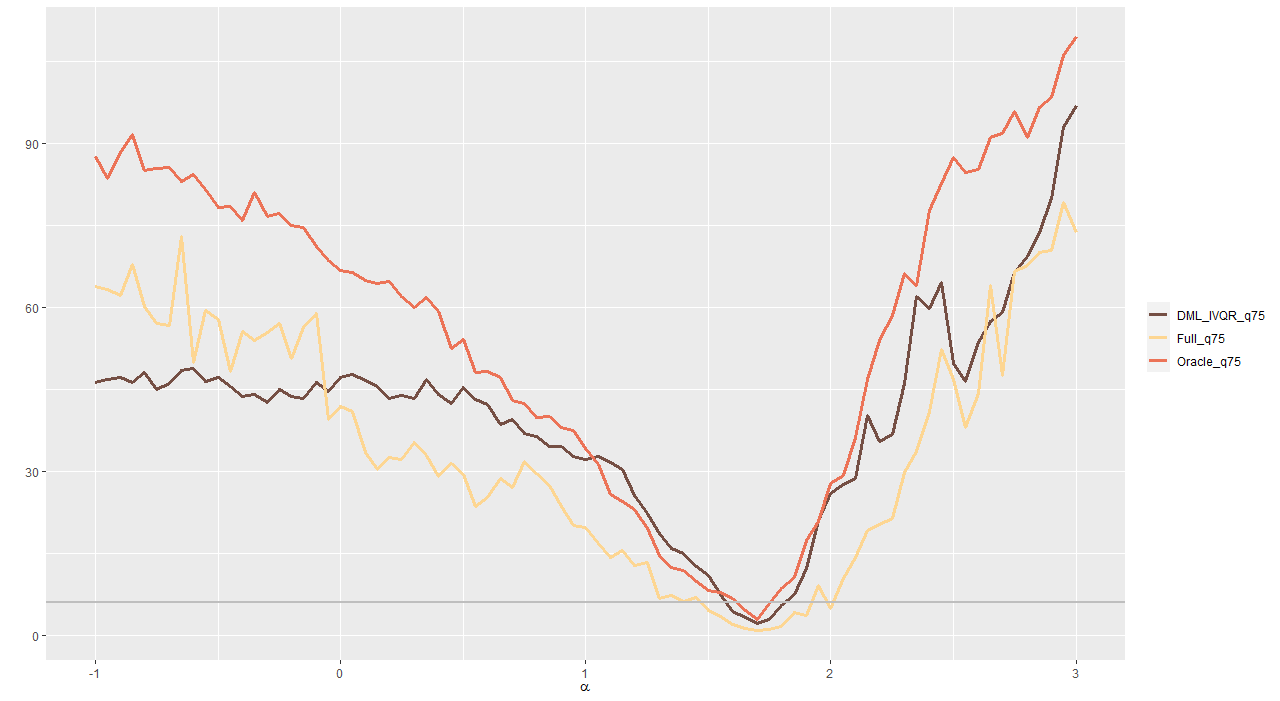}

\end{figure}

\begin{table}[h]
  \caption{Choice of $\lambda$:  \ DML-IVQR with High-dimensional Controls}
  \begin{center}
  \begin{threeparttable}
    \begin{tabular}{lcccccc}
      \hline
           & \multicolumn{3}{c}{n = 500} & \multicolumn{3}{c}{n = 1000}\\
      \hline
            
          & \multicolumn{1}{c}{RMSE} & \multicolumn{1}{c}{MAE} & \multicolumn{1}{c}{BIAS} & \multicolumn{1}{c}{RMSE} & \multicolumn{1}{c}{MAE} &  \multicolumn{1}{c}{BIAS} \\
        \hline
      $\alpha_{0.25}$ ($\lambda=$ Belloni and Chernozhukov) & 0.1716   &  0.1325  & -0.0716  &  0.0849 &  0.0683 & 0.0056 \\
      $\alpha_{0.25}$ ($\lambda=$ 5-fold Cross-Validation) & 0.1720  & 0.1368 & -0.0986 & 0.0995  & 0.0811  & -0.0589 \\
      \hline
      $\alpha_{0.50}$ ($\lambda=$ Belloni and Chernozhukov)   & 0.1273   &  0.0962  & 0.0270 & 0.0800 & 0.0556  & 0.0384 \\
      $\alpha_{0.50}$ ($\lambda=$ 5-fold Cross-Validation) & 0.1374  & 0.1032 &  -0.0384  & 0.0779  & 0.0536   & -0.0236 \\
      \hline
      $\alpha_{0.75}$ ($\lambda=$ Belloni and Chernozhukov)  & 0.1572  & 0.1272  & 0.0876  & 0.1142  & 0.0961 & 0.0839 \\
      $\alpha_{0.75}$ ($\lambda=$ 5-fold Cross-Validation) & 0.1526 & 0.1179 & 0.0286 & 0.0838  & 0.0677 & 0.0205 \\
      \hline
      \end{tabular}%
    
    \label{table 3}%
    
\begin{tablenotes}
  \tiny
    \end{tablenotes}
  \end{threeparttable}
  \end{center}
  \label{table: lambda}
  \end{table}

\section{An Empirical Study: Quantile Treatment Effects of 401(k) Participation on Accumulated Wealth}
In this section, 
we reinvestigate an empirical study on quantile treatment effects: 
the effect of 401(k) participation on wealth, cf.\ \cite{ref-ChernozhukovHansen2004}.
Not only does this conduct data-driven robustness checks on the econometric results, 
but the DML-IVQR sheds light on the treatment effect heterogeneity among the control variables. 
This complements the existing empirical findings. 
In addition, we compare our empirical results with those from \cite{ref-ChenHsiang2019} that conduct
the IVQR estimation by using generalized random forest approach,
which is an alternative in causal machine learning literature.

Examining the effects of 401(k) plans on accumulated wealth is an issue of long-standing empirical interest. 
For example, based on the identification of selection on observables, 
\cite{ref-ChiouChenChen2018} and \cite{ref-ChernozhukovHansen2013} suggest that 
the income nonlinear effect exists in the 401(k) study. 
Nonlinear effects from other control variables are identified as well.

Based on DML-IVQR, we reinvestigate the impact of the 401(k) participation on accumulated wealth.
Total wealth (TW) or net financial assets (NFTA) is the outcome variable $Y$.
The treatment variable $D$ is a binary variable that stands for participation in the 401(k) plan.
Instrument $Z$ is an indicator of eligibility to enroll in the 401(k) plan.
The vector of covariates $X$ consists of income, age, family size, marriage, an IRA individual retirement account, a defined benefit status indicator, a home ownership indicator and the different education-year indicator variables. 
The data consists of 9915 observations.

Following the regression specification set up in \cite{ref-ChernozhukovHansen2004},
Table 3 presents quantile treatment effects obtained from different estimation procedures which 
have been defined in the previous sections including
IVQR, res-GMM and GMM. 
The resulting estimates are similar.
As to the high-dimensional analysis, we create 119 technical control variables 
including those constructed by polynomial bases, interaction terms, and cubic splines (thresholds). 
To ensure each basis has equal length, 
we utilize the minimax normalization for all technical control variables.
Consequently, we use the plug-in method to determine the penalty value when performing the Lasso under the moment condition, and tune the penalty in the quantile $L1$-norm objective function based on
the Huber approximation by 5-fold cross-validation.
The DML-IVQR also implements feature normalization of the outcome variable for  computational efficiency. 
To make the estimated treatment effects across different estimation procedures roughly comparable, 
Table 5 shows that the effect obtained through the DML-IVQR is multiplied by the standard deviation of the outcome variable.
Weak identification/instrument robust inference on quantile treatment effects are depicted in 
Figures 5 and 6.
However, the robust confidence interval widens as the sample size decreases at the upper quantiles.
Estimated quantile treatment effects are significantly different from zero.
We can use the result from the DML-IVQR as a data-driven robustness check on 
those summarized in the Table 4.

Tables 6 and 7 present the selected important variables across different quantiles.
The approximate sparsity is asymmetric across the conditional distribution in the sense that
the number of selected variables decreases as the quantile index $\tau$ increases,
although it hinges on a relatively small number of observations at the upper quantiles.
In this particular example, 
$\tau$ captures the rank variable that governs the unobservable heterogeneity: savings propensity.
Small values of $\tau$  represent participants with low savings propensity.
Our empirical results thus signify that the 401(k) participants with low savings propensity are more associated with the nonlinear income effect than those with high savings propensity, which complements the results concluded in previous studies
\cite{ref-Chernozhukov2018} and \cite{ref-ChiouChenChen2018}.
The nonlinear income effects, across quantiles ranging from (0, 0.5], are picked up by the selected variables, such as
$\max(0, inc-0.2)$, $\max(0, inc^2-0.2)$,$\max(0, inc^3-0.2)$ and etc.
Technical variables in terms of age, education, family size, and income are more frequently selected in Tables 6 and 7.
In addition, these four variables are also identified as important variables in the 
context of the generalized random forests, cf.\  \cite{ref-ChenHsiang2019}.

\begin{table}
  \centering
  \caption{Estimations with the Model Specification as in Chernozhukov and Hansen (2004)}

\medskip
  
  \begin{threeparttable}
    \begin{tabular}{l*{6}{c}r}
    \hline
    \hline
    Quantiles             &0.1 &0.15 &0.25 &0.5  &0.75  &0.85 &0.9  \\
    \hline
    \hline
    
    TW(IVQR)  & 4400 & 5300 & 4900 & 6700 & 8000 & 8300 & 10800\\
    \hline
    TW(res-GMM) & 4400 & 5100 & 4900 & 6300 & 8200 & 7500 & 9100\\
    \hline
    TW(GMM) & 4400 & 5200 & 4800 & 6300 & 8400 & 8000 & 8700\\
    \hline

    NFTA(IVQR)  & 3600 & 3600 & 3700 & 5700 & 13200 & 15800 & 17700\\
    \hline
    NFTA(res-GMM) & 3500 & 3600 & 3700 & 5600 & 13900 & 15800 & 17700\\
    \hline
    NFTA(GMM) & 3500 & 3600 & 3700 & 5700 & 13900 & 16100 & 18200\\
    \hline
    \hline
    
    \end{tabular}%
    \begin{tablenotes}
      \scriptsize
     
      \end{tablenotes}
      \end{threeparttable}
  
  \label{table 3}%
\end{table}

\begin{table}
  \centering
  \caption{DML-IVQR with High-dimensional Controls}

  \medskip
  \begin{threeparttable}
    \begin{tabular}{l*{6}{c}r}
    \hline
    \hline
    Quantiles             & 0.1 & 0.15 & 0.25  & 0.5  & 0.75 & 0.85 & 0.9\\
    \hline
    NFTA(std-DML-IVQR $\times$63522) & 3176  & 3049 & 3303 & 5844 & 18802 & 26298 & 28076\\
    \hline
    TW(std-DML-IVQR $\times$111529)  & 2453  &3011 & 3457 &  7695 &  15056 & 18736 & 16394\\
    \hline
    NFTA(std-DML-IVQR) & 0.05  &0.048 & 0.052 & 0.092 & 0.296 & 0.414 & 0.442\\
    \hline
    TW(std-DML-IVQR)  & 0.022  &0.027 & 0.031 & 0.069 & 0.135 & 0.168 & 0.147\\
    \hline
    \hline
    \end{tabular}%
    \begin{tablenotes}
      \scriptsize
We create 119 technical control variables including those constructed by the polynomial bases, 
interaction terms, and cubic splines (thresholds). DML-IVQR estimates the distributional effect
which signifies an asymmetric pattern similar to that identified in Chernozhukov and Hansen (2004).
      \end{tablenotes}
      \end{threeparttable}
  \label{table 4}%
\end{table}

\begin{figure}

  \centering

  \label{figur}\caption{DML-IVQR Weak-Instrument Robust Inference: 401(K) participation on TW}
  
  \bigskip
  \begin{threeparttable}
  \begin{tabular}{cc}


    \includegraphics[width=68mm]{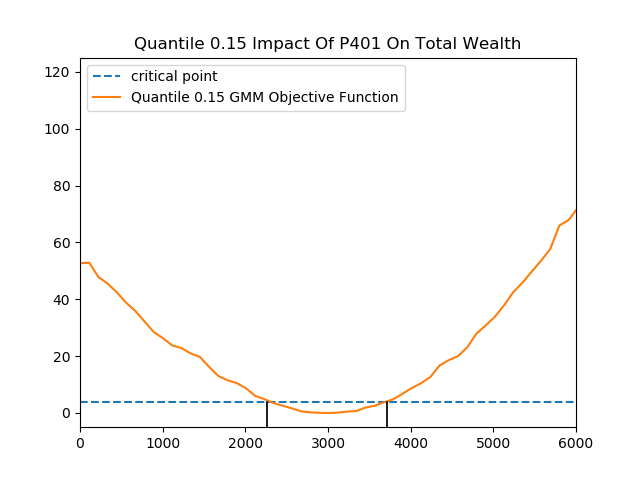}&

    \includegraphics[width=68mm]{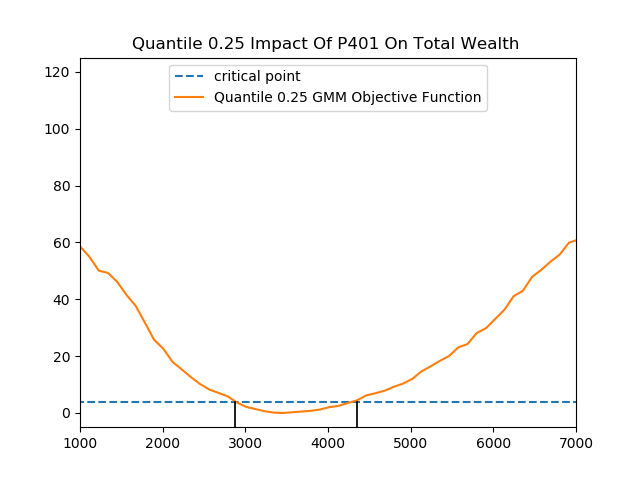}\\

    \includegraphics[width=68mm]{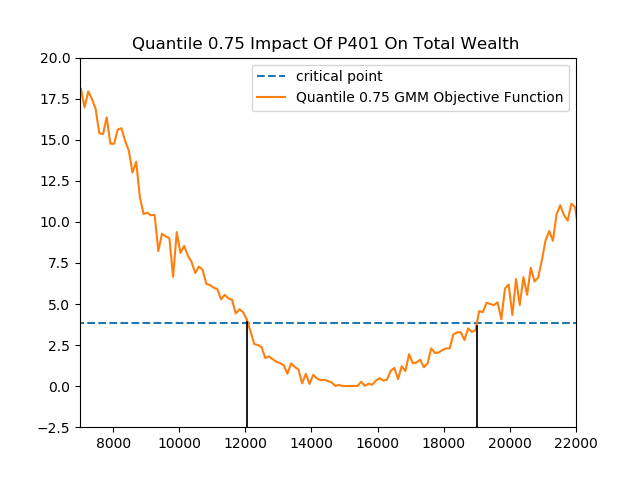}&

    \includegraphics[width=68mm]{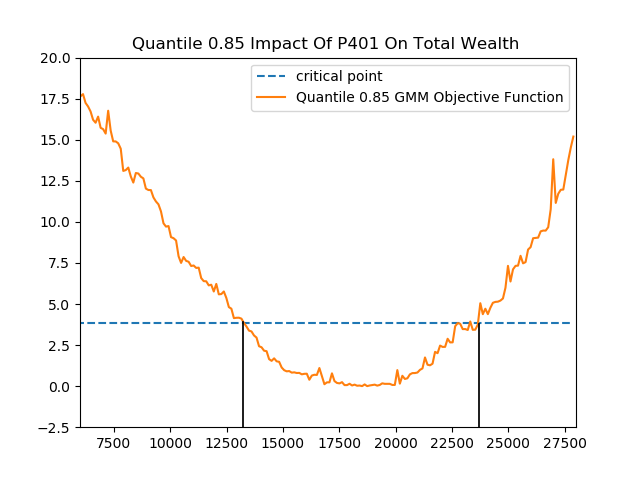}\\

  \end{tabular}
  \begin{center}
  \includegraphics[width=68mm]{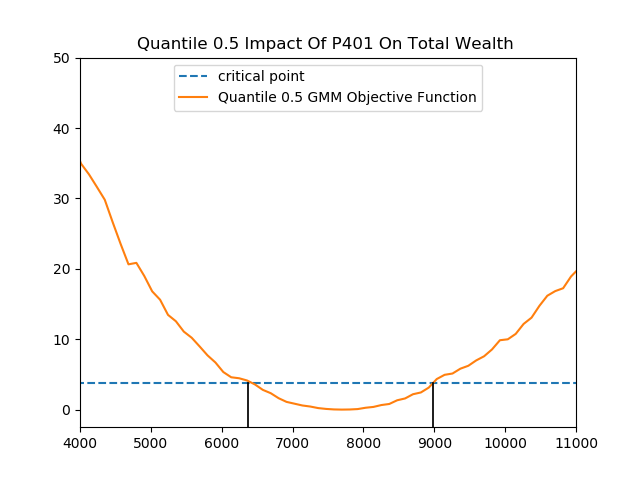}
  \end{center}
 
  \begin{tablenotes}
    \scriptsize
    
    \end{tablenotes}
  \end{threeparttable}
\end{figure}

\begin{figure}

  \centering

  \label{figur}\caption{DML-IVQR Weak-Instrument Robust Inference: 401(K) participation on NFTA}
  
  \bigskip
  \begin{threeparttable}
  \begin{tabular}{cc}


    \includegraphics[width=68mm]{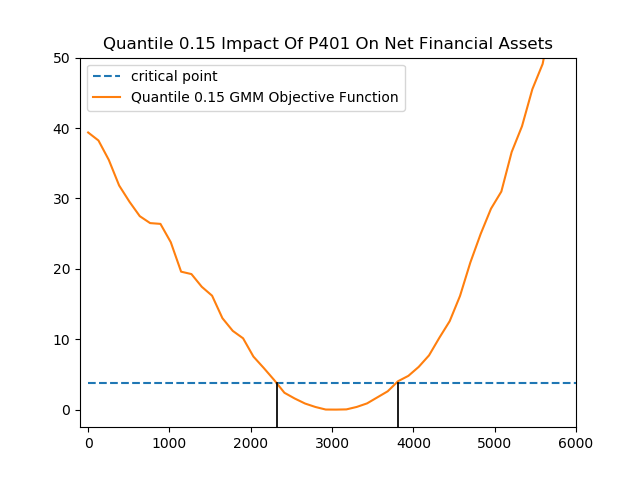}&

    \includegraphics[width=68mm]{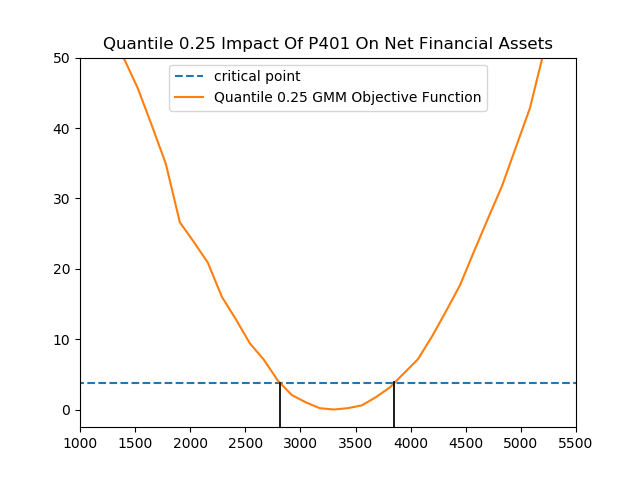}\\

    \includegraphics[width=68mm]{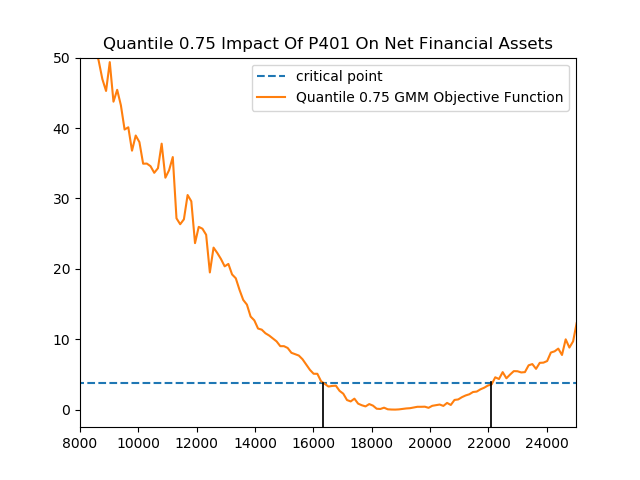}&

    \includegraphics[width=68mm]{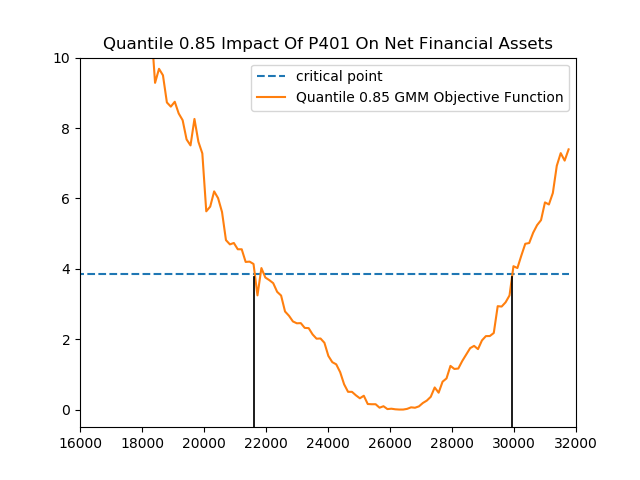}\\

  \end{tabular}
  \begin{center}
    \includegraphics[width=68mm]{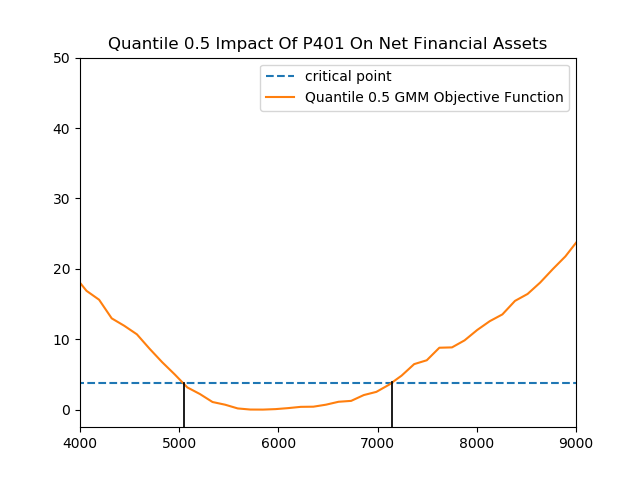}
  \end{center}
  
\begin{tablenotes}
  \scriptsize

   \end{tablenotes}
\end{threeparttable}
\end{figure}

\begin{table}
  \centering
  \caption{Total Wealth}

 \par
  \medskip
    \begin{threeparttable}
  
    \begin{tabular}{|l|{c}|}
    \hline
    \hline
    Quantile            &Selected Variables  \\
    \hline
    \hline
    0.15  & $ira, educ, educ^2, age\times ira, age\times inc, fsize\times educ, fsize\times hmort$ \\ 
        & $ira\times educ, ira\times inc, hval\times inc,  marr, male, i4 , a3$\\ 
        & $twoearn, marr\times fsize, pira\times inc, \max(0, age^3-0.2)$\\ 
        & $\max(0, educ^2-0.4), \max(0, educ-0.2), \max(0, age^2-0.4)$\\
        
    \hline
    0.25  & $ira,  age\times fsize, age\times ira, age\times inc$\\ 
        & $fsize\times educ, ira\times educ, ira\times inc$\\ 
        & $hval\times inc, marr, male, i3, twoearn, marr\times fsize$\\
        & $pira\times inc, twoearn\times fsize, \max(0,inc-0.2)$ \\

    \hline
    0.5  & $inc^2, age\times fsize, age\times ira, age\times inc$\\ 
        & $fsize\times educ, ira\times educ, ira\times hval, ira\times inc$\\ 
        & $hval\times inc, male, a1, a3 , pira\times inc, twoearn\times age, twoearn\times fsize$\\
        & $twoearn\times hmort, twoearn\times educ, \max(0, educ-0.6)$ \\
        
    \hline
    0.75  & $inc, ira,  age\times ira,  age\times hval$\\ 
        & $age\times inc, educ\times inc, hval\times inc, pira\times inc, pira\times age$\\
        
    \hline
    0.85  & $inc, ira, age\times hval, age\times inc, ira\times educ$\\ 
        & $educ\times inc, hval\times inc, pira\times inc, pira\times hval$ \\ 
    \hline
    \hline
    \end{tabular}%
    \begin{tablenotes}
      \scriptsize
     Selected variables across $\tau$, tuned via cross validation.
      \end{tablenotes}
      \end{threeparttable}
  \label{table 5}%
\end{table}

\begin{table}
  \centering
  \caption{Net Financial Assets}


  \par
  \medskip
    \begin{threeparttable}
    \begin{tabular}{|l|{c}|}
    \hline
    \hline
    Quantile            &Selected Variables \\
    \hline
    \hline
    0.15  & $ira, educ\times 2, fsize\times 3, hval\times 3, educ\times 3, age\times educ, age\times hmort$\\ 
        &  $age\times inc, fsize\times hmort, fsize\times inc, ira\times educ, ira\times inc$\\ 
        & $hval\times inc, marr, db, male, i2, i3$\\ 
        & $i4,i5, twoearn, marr\times fsize, pira\times inc, pira\times educ, twoearn\times inc$\\
        & $twoearn\times ira,\max(0,age^3-0.2), \max(0,age^2-0.2), \max(0,age-0.6)$\\
        & $\max(0,inc^3-0.2), \max(0,inc^2-0.2), \max(0,educ-0.2)$
 \\
    \hline
    0.25  & $ira, hmort, age\times hmort, age\times inc, fsize\times hmort, fsize\times inc$\\ 
        & $ira\times educ, ira\times inc, hval\times inc,db, smcol, male$\\ 
        & $i2, i3, i4, i5, a2, a3, twoearn, pira\times inc, pira\times age$\\
        & $pira\times fsize, twoearn\times inc,  twoearn\times ira, twoearn\times hmort, max(0, age^2-0.2)$ \\
        & $\max(0, age-0.6), \max(0, inc^2-0.2), \max(0, inc-0.4)$ \\
        & $\max(0, inc-0.2), \max(0, educ-0.2)$
 \\

    \hline
    0.5  & $age, ira, age\times fsize,  age\times ira, age\times inc$ \\ 
        & $fsize\times educ ,   fsize\times hmort, ira\times educ,  ira\times inc, hval\times inc, hown$\\ 
        & $male, i3, i4, a1, a2, a4, pira\times inc, pira\times fsize, twoearn\times inc, twoearn\times fsize$\\
        & $twoearn\times hmort, twoearn\times educ, \max(0,inc-0.2)$
 \\
    \hline
    0.75  & $ira, age\times inc, hval\times inc, pira\times inc, pira\times age$
     
 \\
    \hline
    0.85  & $ira,  age\times inc,  educ\times inc,   hval\times inc, pira\times inc$\\ 
    
    \hline
    \hline
    \end{tabular}%
    \begin{tablenotes}
      \scriptsize
      Selected variables across $\tau$, tuned via cross validation.
      \end{tablenotes}
      \end{threeparttable}
  \label{table 5}%
\end{table}

\section{Conclusion}

In this study, we investigate the performance of a debiased/double machine learning algorithm within the framework of high-dimensional IVQR. The simulation results indicate that our procedure performs more efficiently than those based on conventional estimators with many controls. Furthermore, we evaluate the corresponding weak identification robust confidence interval of the low-dimensional causal parameter. 
Given many technical controls, we reinvestigate  thequantile treatment effects of the 401(k) participation on accumulated wealth and then highlight the non-linear income effects across the savings propensity.

\vspace{6pt} 



\authorcontributions{Three authors contributed equally to the paper.}

\funding{This research was funded by the JSPS KAKENHI (Grant no. JP20K01593), 
the personal research fund from Tokyo International University, and it was
financially supported by the Center for Research in Econometric Theory and Applications (Grant no. 107L900203 ) from The Featured Areas Research Center Program within the framework of the Higher Education Sprout Project by the Ministry of Education (MOE) in Taiwan.
}

\acknowledgments{We are grateful to Tsung-Chih Lai and Hsin-Yi Lin for discussions and comments. This paper has benefited from presentations at Ryukoku University and the 2nd International Conference on Econometrics and Statistics (EcoSta 2018). The usual disclaimer applies.}

\conflictsofinterest{The authors declare no conflict of interest.} 

\abbreviations{The following abbreviations are used in this manuscript:\\

\noindent 
\begin{tabular}{@{}ll}
DML & Double machine learning\\
GMM & Generalized method of moments\\
GRF & Generalized random forests\\
IVQR & Instrumental variable quantile regression\\
Lasso & Least absolute shrinkage and selection operator
\end{tabular}}

\appendixtitles{no} 

\reftitle{References}





\end{document}